\documentclass[wcp,twocolumn,10pt]{jmlr} 

\usepackage{booktabs}
\usepackage{siunitx}
\usepackage{authblk}
\usepackage[switch]{lineno}

\theorembodyfont{\upshape}
\theoremheaderfont{\scshape}
\theorempostheader{:}
\theoremsep{\newline}

\title[Explainable Machine Learning Models in Medical Image Analysis]{Exploring Radiologists' Expectations of Explainable Machine Learning Models in Medical Image Analysis}

\author[1,2,3]{Sara Ketabi}
\author[1,4]{Matthias W. Wagner}
\author[1,2]{Birgit Betina Ertl-Wagner}
\author[2]{Greg A.Jamieson}
\author[1,2,3]{Farzad Khalvati}
\affil[1]{The Hospital for Sick Children}
\affil[2]{University of Toronto, Toronto, Canada}
\affil[3]{Vector Institute for Artificial Intelligence, Toronto, Canada}
\affil[4]{University Hospital Augsburg, Augsburg, Germany}

\affil[*]{farzad.khalvati@utoronto.ca}

\begin{document}

\maketitle

\begin{abstract}
In spite of the strong performance of machine learning (ML) models in radiology, they have not been widely accepted by radiologists, limiting clinical integration. A key reason is the lack of explainability, which ensures that model predictions are understandable and verifiable by clinicians. Several methods and tools have been proposed to improve explainability, but most reflect developers’ perspectives and lack systematic clinical validation. In this work, we gathered insights from radiologists with varying experience and specialties into explainable ML requirements through a structured questionnaire. They also highlighted key clinical tasks where ML could be most beneficial and how it might be deployed. Based on their input, we propose guidelines for designing and developing explainable ML models in radiology. These guidelines can help researchers develop clinically useful models, facilitating integration into radiology practice as a supportive tool.
\end{abstract}

\paragraph*{Data and Code Availability}
The survey data and analysis code used in this study will be released upon paper acceptance.

\paragraph*{Institutional Review Board (IRB)}
The IRB information will be provided if the paper is
accepted.

\section{Introduction}
\label{intro}
Machine learning (ML) models have shown considerable potential in a wide range of clinical tasks in radiology, such as image classification \citep{esteva2017dermatologist}, image segmentation \citep{havaei2017brain}, and radiology report generation \citep{wang2023r2gengpt}, with some models even performing on par with radiologists. Nevertheless, these models have not been widely validated by radiologists, resulting in their limited adoption in real-world clinical procedures. One of the major reasons for this is that radiologists usually struggle to understand and validate model output, which is referred to as lack of explainability. Explainability is an essential characteristic of ML models which ensures that radiologists can be in the loop of model decisions, thereby encouraging them to integrate these models into their clinical workflow \citep{arrieta2020explainable}. Explainability is one of several factors contributing to the broader concept of ``trust" in ML models, alongside others including performance, confidence, and fairness. 

Several works in the literature have considered explainability as an important element of ML models, applying different methods, such as self-explainable approaches and post-hoc explanations, for improving it. Self-explainable approaches include simple conventional ML models, e.g., decision trees, or Deep Learning (DL) models that involve additional components integrated into the model during training to make it explainable by design, e.g., trainable attention mechanism \citep{vaswani2017attention} or prototypical networks \citep{snell2017prototypical}. Post-hoc explanations, on the other hand, refer to model-agnostic methods applied to the model after training to reveal the information used for generating an outcome, such as Gradient-Weighted Class Activation Mapping (GradCAM) \citep{selvaraju2017grad} or local interpretable model-agnostic explanations \citep{ribeiro2016should}.



Most of these studies, however, investigate explainability from a technical perspective, without involving clinicians to validate their proposed methods. This leads to a gap between what these studies provide as explanations and what radiologists say they need to verify model decisions. To fill this gap, several works in the literature e.g., \citep{tonekaboni2019clinicians} have conducted organized surveys to obtain clinicians' feedback regarding model explainability, which provided us with useful insights into designing the survey used in our study. Nonetheless, the specific challenges existing in radiology, such as Complex and multi-step diagnostic workflows, requires significant consideration in the design of ML models applied to this field. These considerations can be addressed through conducting systematic evaluation methods targeted at radiologists, which is the main contribution of this paper.

In this study, we surveyed a group of radiologists to obtain their perspective on explainable ML models through a structured questionnaire. Analyzing their responses will enable us to get detailed insights into the expectations of specific radiological subgroups. We also asked them to mention potential areas where these models can be useful in improving radiologists' efficiency in clinical tasks as well as crucial model evaluation aspects. Based on this analysis and the gaps identified through the survey findings, we will propose a set of guidelines containing necessary elements required by radiologists for improving model explainability, helping them validate model decisions more efficiently.



\section{Related Work}

\subsection{Explainability Methods for Medical Image Analysis}
\begin{figure*}[t]
\centering
\includegraphics[width=3.5in]{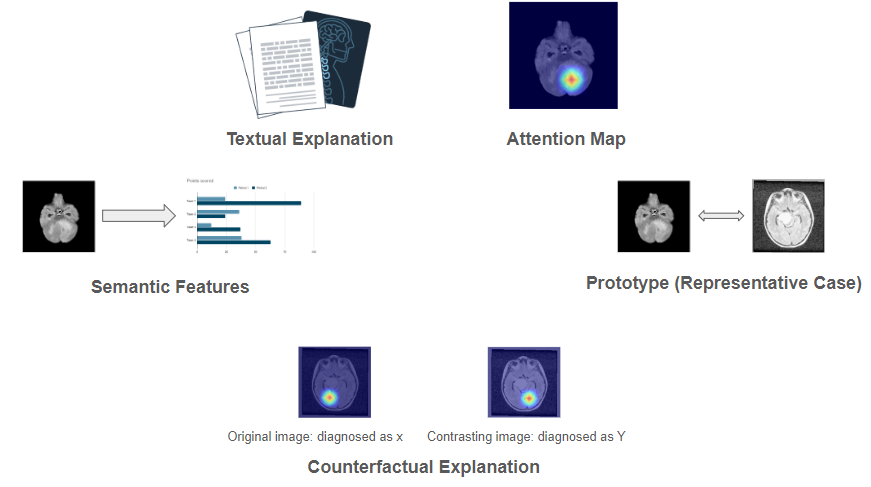} 
\caption{Schematic Overview of Explainability Methods for Medical Image Analysis}
\label{xai}
\end{figure*}

Prior works have identified explainability as an essential element of ML models, particularly for their acceptance and adoption in clinical settings, and proposed various approaches to enhance it. These approaches can be divided into two main categories, namely self-explainable and post-hoc. Self-explainable methods integrate explainability elements into model training, making the model interpretable by design \citep{hou2024self}. These methods include, but are not limited to, attention-based, prototype-based, counterfactual explanations, concept-based (semantic features), and textual reasoning \citep{hou2024self}. Figure \ref{xai} provides a brief schematic representation of these methods.

Attention-based methods apply trainable attention weights \citep{vaswani2017attention} to provide visual heat maps that indicate parts of an input image attended by the model when producing a certain prediction. Concept-based methods provide explainability by relating final model predictions to a set of high-level clinical concepts. These concepts can be in the form of textual semantic features \citep{salahuddin2022transparency} or standardized radiologic terminology and classification systems (SRTCS), which refers to radiologist-provided scores used for interpreting a diagnostic task, e.g., the breast imaging reporting and data system (BIRADS) \citep{ozsahin2024bi}. Prototype-based methods \citep{snell2017prototypical} illustrate a model's decision for a certain case by comparing it with a set of predetermined prototypes (representative cases) \citep{hou2024self}. Textual explanations refer to additional elements in the form of natural language provided alongside model predictions to support them, e.g., using a language transformer  \citep{hou2024self,chen2020generating}. Counterfactual explanations work by applying minimal changes to imaging features to modify the corresponding original output, creating a contrastive sample \citep{hou2024self}.

Post-hoc methods, on the other hand, refer to model-agnostic algorithms applied after model training to indicate the contribution of input features to a specific prediction. These methods can be divided into gradient-based and perturbation-based subcategories. Gradient-based methods specify important features by taking the output gradient with respect to the input, such as Gradient-weighted Class Activation Mapping (GradCAM) \citep{selvaraju2017grad}. Perturbation-based methods work by changing different parts of the input data and using the resulting effect on the final prediction to determine feature contribution \citep{singh2020explainable}, e.g., local interpretable model-agnostic explanations \citep{ribeiro2016should}.  

In spite of the capability of the aforementioned methods in highlighting the rationale behind ML models' decisions, their current application to clinical tasks typically lacks radiologists' verification. In other words, many of these state-of-the-art explainability tools have been designed based on ML developers' opinions, making them biased towards developers' subjective perceptions. However, effective deployment of ML models in clinical settings requires extensive user studies to identify the specific expectations of healthcare professionals regarding explainable ML and understand how these explanations can encourage clinical adoption, an area that has received limited attention in previous research.


\subsection{User Studies for Gathering Clinicians' Feedback on Explainable ML Models}

In fact, only a few works in the literature have performed user studies to collect clinicians' point of view on explainable ML in healthcare. In a work performed by \citep{tonekaboni2019clinicians}, 10 clinicians from Intensive Care Unit and Emergency Department interviewed to investigate their understanding of explainability, determine appropriate explanation methods based on their needs, highlight how ML can address these considerations, and propose explainability evaluation metrics. Although some of the findings of this study can be useful in developing imaging-based models, the unique challenges of medical imaging data, such as high dimensionality, subtle diagnostic features, and the need for precise visual explanations, require explicit user studies to identify images-specific explainability elements.

In the field of medical imaging, CheXplain \citep{xie2020chexplain} is a human-centered framework involving a survey from a group of 39 physicians and 38 radiologists analyzing chest X-ray images to understand how these radiologists explain their diagnoses to physicians. This would provide some insights into the requirements of clinicians for ML models for chest X-ray diagnosis. Another work \citep{jin2022evaluating} conducted a survey along with an optional interview with 6 neurosurgeons to get their evaluation of 16 post-hoc explainability methods in the form of heatmaps, e.g., GradCAM, applied to the brain tumor grading task. The evaluation was based on how well the heatmap-specified areas would align with their clinical judgment. A user study \citep{gulum2024explainable} was conducted with 10 radiologists, who were provided with a set of post-hoc and uncertainty-based explanations for prostate lesion detection. The goal was to investigate the effect of using these explanations on the radiologists' diagnostic performance as well as highlighting the most desirable explanation tools among the provided ones through a questionnaire.

Despite the significant impact of these user studies on enhancing the development of explainable ML models for medical image analysis, their design has been limited to a single imaging modality or use case. As a result, their results may not fully generalize to other imaging modalities, where radiologists might have distinct requirements or priorities. To the best of our knowledge, our work is the first radiologist-centered study designed for multiple medical imaging modalities and clinical tasks to collect and analyze radiologists' expectations from ML explainability. Therefore, it offers a practical framework to support model explainability in radiology and facilitate its integration into clinical practice.
\vspace{-0.085in}
\section{Methodology}
In this Quality Improvement-approved study, we prepared a questionnaire containing 30 questions using Microsoft Forms and distributed it among 46 radiologists with various years of experience from different hospitals in a metropolitan area. Twenty eight questions were designed in a multiple-choice format, while the remaining 2 were in free text, allowing the participants to provide as much detail as they would like. In 11 of the multiple-choice questions, we also included an ``other" option to enable the radiologists to specify any additional answer of their choice. The average completion time for the survey was 20 minutes per participant. The information related to the participants in this survey is provided in Subsection \ref{participants}, and the detailed description of the questionnaire can be found in Subsection \ref{survey structure}. 


\label{methods}
\subsection{Participants}
\label{participants}
\begin{figure}
\centering
\includegraphics[width=3.5in]{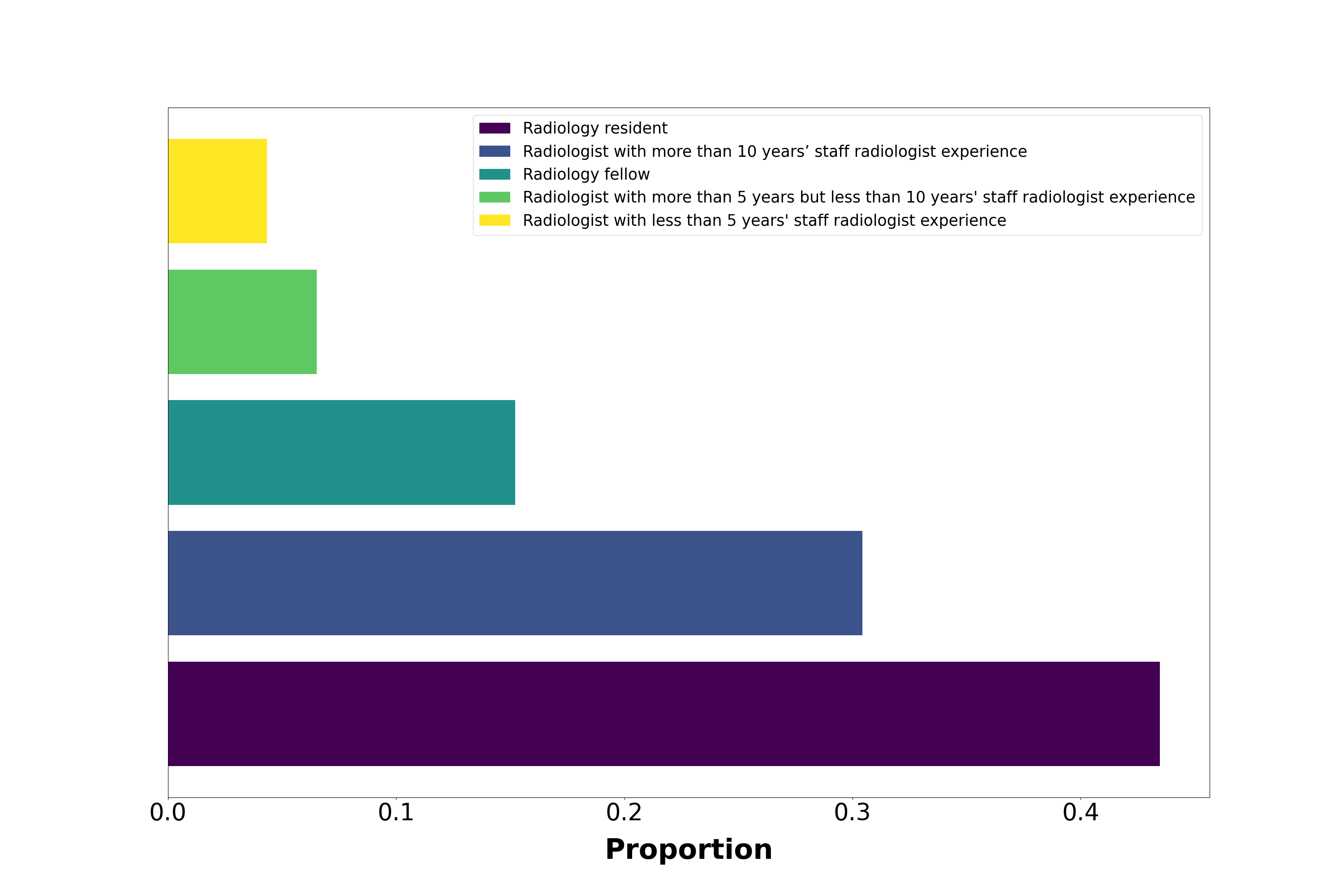} 
\caption{The Radiological Experience of the Survey Participants}
\label{experience}
\end{figure}

At the beginning of the questionnaire, we included a few questions to collect demographic information about the participants. The questionnaire was distributed within the Department of Radiology of the University. Based on the collected responses, there were a total of 26 radiologists and 20 radiology trainees (residents) from different hospitals of the university participating in this survey. Figure \ref{experience} indicates the distribution of the radiological experience of the participants. As can be observed, the majority of the participant were radiology residents, with a distribution of 43\%. Moreover, senior radiologists with more than 10 years of experience constitute the second-largest proportion, i.e., 30\%. This demonstrates that a high number of responses are from the least and most experienced groups, helping us identify the range of expectations from ML models based on the experience of their end-users.

Figures \ref{focus} and \ref{modality} in Appendix \ref{appendix2} demonstrate the proportion of the participants' main clinical/radiological areas and the imaging modalities used in these areas, respectively. The majority of the radiologists were from Abdominal and Neuroimaging divisions. Furthermore, Computed tomography (CT) and Magnetic Resonance Imaging (MRI) were the most frequent imaging modalities in the primary focus of these radiologists.

Additionally, we gathered information about the familiarity of the respondents with data and statistical analysis as well as general ML concepts. These are depicted in Figures \ref{stat} and \ref{ML} in Appendix \ref{appendix2}. Most of the radiologists had basic knowledge of statistical analysis, and some of them were experienced with applying these methods to research or clinical practice. Regarding the ML experience, 52\% of the radiologists did not have significant experience with ML, with 26\% of them knowing ML concepts at a basic level. Furthermore, only a few of them had applied ML to research or clinical practice. Therefore, most of these radiologists were not familiar with the detailed functionality of ML models and how these models can make a prediction on a certain dataset.

\subsection{Survey Structure}

Our questionnaire consists of 30 questions in 9 sections. Sections 1 and 2 collect general information about the participants, including their radiological experience, clinical-radiological primary focus area, and the imaging modality used in this area. In Section 3, the radiologists would answer two questions to indicate their level of familiarity with statistical, data analysis, and ML concepts. Subsequently, in Section 4, we asked the participants to mention the clinical task(s) in which ML models can be useful for clinical end use. 

Our main explainability-related questions begin in Section 5, where radiologists were asked whether they consult clinical information from cases similar to the one at hand to improve their decision-making. Moreover, we inquired about the tools radiologists employ to explain their diagnosis to their colleagues or physicians, including imaging features and information from the internet. This enables us to identify the key sources of information they rely on to confirm a diagnosis, insights that can also help validate the outputs of diagnostic ML models.

The goal of Section 6 of the questionnaire is to recognize the elements making an ML model applied to a medical image classification task more explainable to radiologists, enabling them to validate model predictions and integrate into their clinical practice. First, we asked them why they want an ML model to be explainable. This would highlight how radiologists  could benefit from explainability approaches, e.g., by understanding the features contributing to a model's prediction, better justifying a certain diagnosis to a colleague, and getting confidence in writing radiology reports. Furthermore, we included a question regarding regions of interest (ROIs)-agnostic models to see whether it is important for an ML model to be trained on ROIs, e.g., abnormality-related areas specified by radiologists, to encourage them to integrate the model into their workflow. As a key question, we also requested the participants to highlight the element(s) making an ML model explainable to them, e.g., basing predictions on clinically relevant imaging features. 

 Section 7 of the questionnaire contains a set of questions aiming to determine whether specific considerations in model development process, from data preprocessing to performance evaluation, can make the model explainable and/or trustworthy to radiologists.  These include, but are not limited to, transparent data preprocessing, transparent model architecture, and involving a feedback mechanism to enable the end-user to adjust model predictions. These insights can determine whether addressing these considerations at different stages of model development improves radiologists' understanding and/or trust in the model's outputs.

Section 8 of the questionnaire is designed to identify important model validation requirements from radiologists' perspective. More specifically, we aimed to determine whether model performance, as it directly impacts clinical decision accuracy and safety, is the main aspect radiologists look into when evaluating an ML model, or if they examine other elements including model explanations and confidence measurements. Moreover, we included a challenging question to find out what radiologists would do in situations where model output contradicts their diagnosis. This would investigate the impact of model explanations on adjusting or changing radiologist's decision. Additionally, we sought to identify the necessity of different model validation types, such as multi-site and prospective validation.

Finally, in Section 9 of the questionnaire, we requested the participants to specify model usability considerations. To that end, we first included a question to detect their major concern regarding the integration of ML models into the clinical workflow, e.g., model performance, lack of explainability, or the possibility of ML replacing them in their roles. In other questions, we investigated the participant's overall interest in using ML models in clinical practices and whether it is necessary for these models to be integrated into a frequently used software by radiologists, e.g., Picture Archiving and Communication System (PACS), prior to clinical deployment. At the end, we provided open-ended fields for the participants to specify particular areas where ML models can significantly enhance the efficiency of their workflow, including those they are uncertain about.

Table \ref{tab1} shows a summary of the information mentioned in this subsection, highlighting a brief description of the questions included in each section of our questionnaire.

\begin{table*}
    \centering
    \caption{Description of the Different Sections Included in Our Questionnaire}
    \vspace{1mm}
    \begin{tabular}{|l|l|l|}
        \hline
        \textbf{Section} &
        \textbf{Topic} \\
        \hline
        1 & Participants' Radiological Experience\\
        
        \hline
        2 & Participants' Radiological Specialty  
 \\
        \hline
        
        3  & Participants' Knowledge and Experience with Statistics and ML\\
        \hline
        4  & Clinical Tasks\\
        \hline
        5  & Radiologists' Diagnostic Workflow \\
        \hline
        6  & Explainability Requirements\\
        \hline
        7  & Explainability and Trust\\
        \hline
        8  & Model Validation\\
        \hline
        9  & Potential Model Usability\\
        \hline
        
    \end{tabular}
    \vspace{5pt}
    
    \label{tab1}
\end{table*}
 
\label{survey structure}

\section{Results} 
\label{results}
In this section, we analyze the participants' responses to our questionnaire. To that end, we first include the analysis in Subsections \ref{4.1}, \ref{4.2}, \ref{4.3}, and \ref{4.4} in detail, where each subsection provides a different perspective on the results. For essential questions, particularly those aimed at understanding the fundamental aspects of model explainability, we filter the responses based on the radiologists' experience, providing separate illustrations for senior radiologists and residents' responses. This ensures that the responses accurately reflect the unique experiences and practices of the participants. We use this analysis to first explain the existing gaps in the current explainable ML literature in Subsection \ref{4.5}. Then, in Subsection \ref{4.6}, we propose a set of guidelines to fill these gaps to assist ML researchers in developing effective explainable models based on the requirements specified by radiologists, increasing the probability of model deployment.

\subsection{Clinical Use Cases of ML}
\label{4.1}
\begin{figure}
\centering
\includegraphics[width=3.6in]{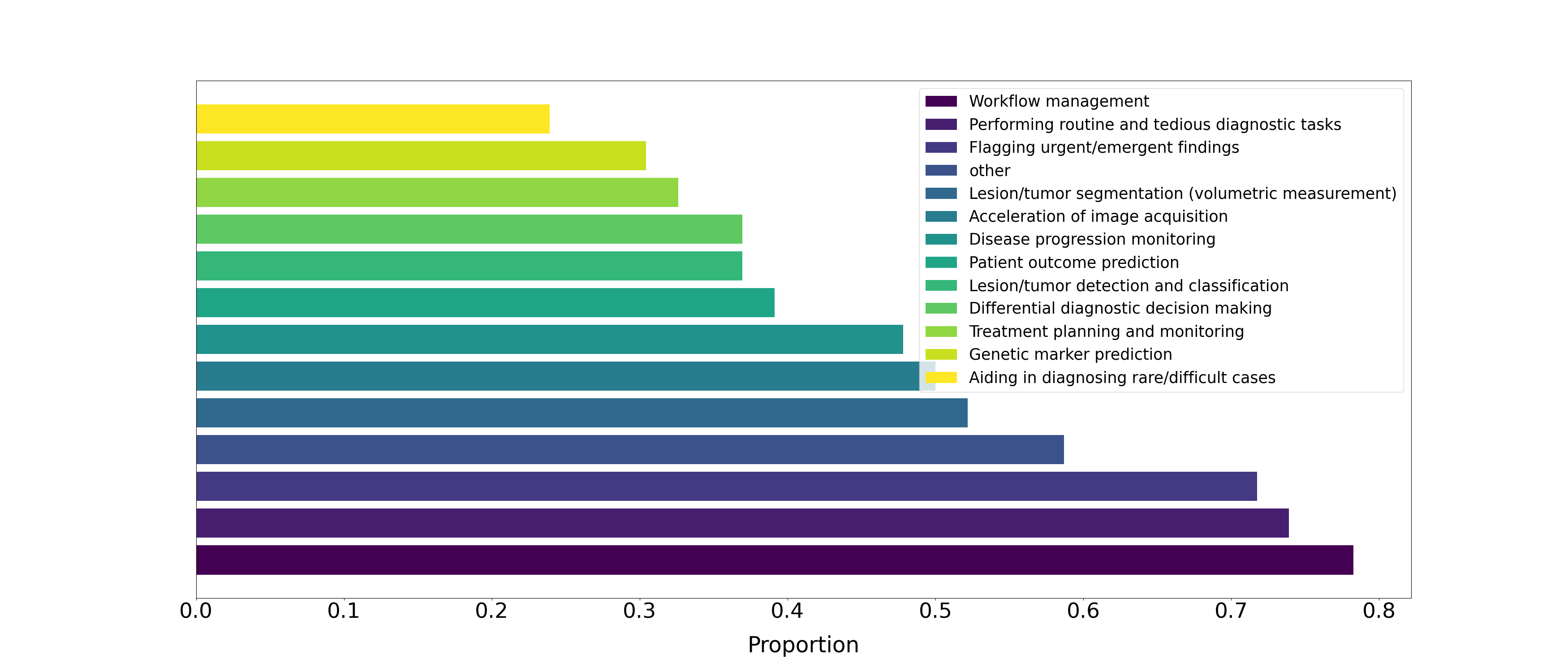} 
\caption{Specific Applications of ML in Clinical Practice}
\label{tasks}
\end{figure}

Among the 46 respondents, 43 indicated their interest in using ML in the clinical workflow. The main specified reasons for this include increasing the efficiency and speed of the clinical workflow. Some of the participants also mentioned that ML can eventually improve their accuracy. Figure \ref{tasks} demonstrates the clinical tasks where ML can be most useful to the participants, highlighted by them. ``Workflow management", ``performing routine and tedious diagnostic tasks", and ``flagging urgent/emergent finding" were the most frequently selected options. On the other hand, very few respondents chose ``aiding in diagnosing rare or difficult cases".

Additionally, exploring the responses to the open-ended questions revealed interesting insights. As shown in Figure \ref{open_ended} in Appendix \ref{appendix2}, enhancing radiologists' efficiency, e.g., in reporting normal cases which are not associated with significant findings, was one of the most common responses (13\%). Other responses include serving as a double-reader in diagnostic tasks, detecting subtle elements missed by radiologists, segmentation, and flagging urgent cases. Furthermore, the participants mentioned that they hope but are unsure whether ML could assist them in differential diagnosis, interpreting rare cases, and report preparation. This is depicted in Figure \ref{open_ended_2} in Appendix \ref{appendix2}.

\subsection{Explainability Requirements}
\label{4.2}
To determine how radiologists perceive explainability, we first analyze the responses related to the tools they use in their clinical workflow to explain their diagnosis to their colleagues. According to Figure \ref{tools} in Appendix \ref{appendix2}, 44 participants (96\%) selected ``Important imaging features" as the most frequent explanation tool. Other frequent responses include ``Information from the Internet" (65\%), ``Original research articles" (59\%), and ``Review articles" (59\%). Moreover, as can be observed in Figure \ref{similar} in Appendix \ref{appendix2}, 18 participants indicated that``examples of similar cases/patients" can ``sometimes" assist them in their decision-making process, with 15 other participants using it ``in most cases".

\begin{figure}[ht!]
\centering
\includegraphics[width=3.5in]{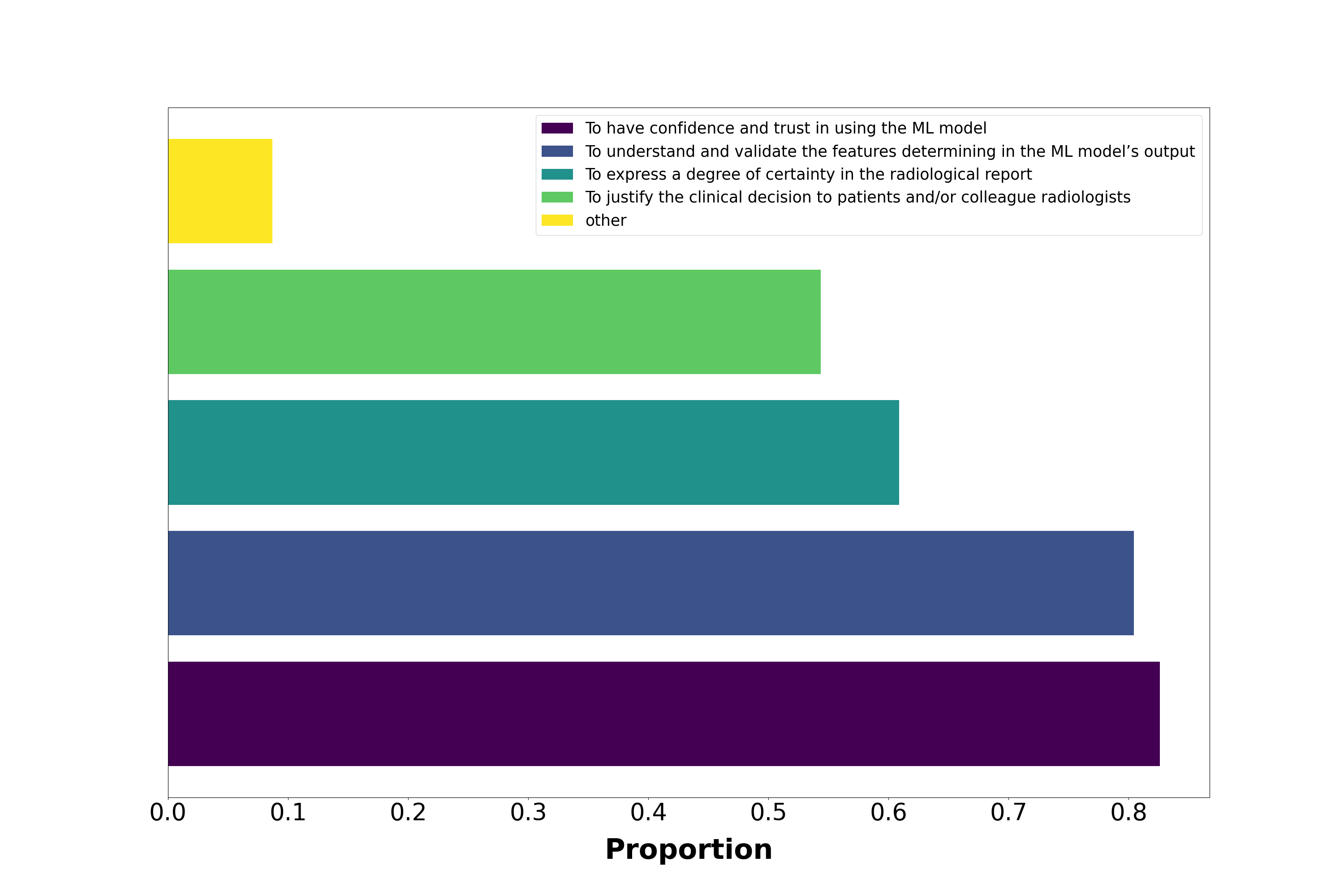} 
\caption{Reasons for Requiring ML Explainability}
\label{reason}
\end{figure}

To understand the detailed requirements of radiologists for ML explainability, we explored the reason that they want an ML model to be explainable, which is represented in Figure \ref{reason}. To that end, most of them indicated ``To have confidence and trust in using the ML model" as the most prominent reason. Moreover, ``To understand and validate the features determining in the ML model’s output" and ``To express a degree of certainty in the radiological report" were the other frequently chosen reasons. 

\begin{figure*}[t]
\centering
\includegraphics[width=5in]{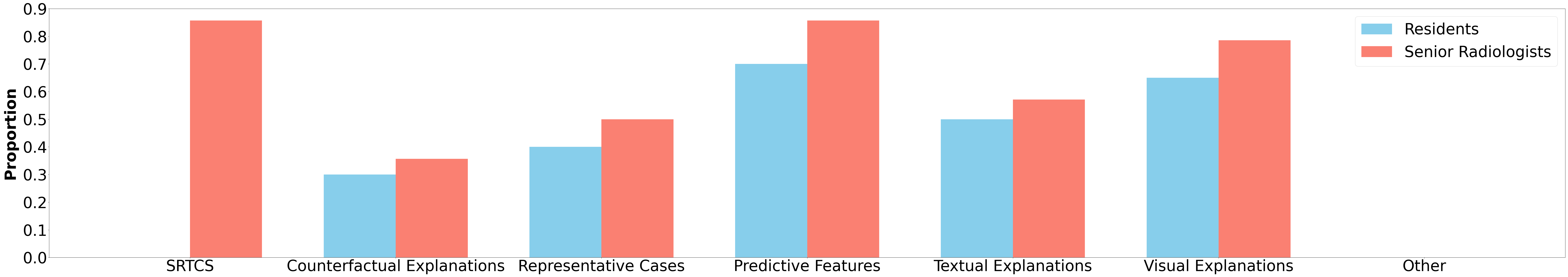} 
\caption{Factors making an ML model explainable to the participants}
\label{factors}
\end{figure*}

According to Figure \ref{factors}, senior radiologists consider a broader range of factors necessary for ML explainability compared to residents. ``Focusing on relevant features, e.g., radiomics (quantitative imaging features), for making a prediction" (residents: 70\%, senior radiologists: 86\%) and ``providing visual heatmaps indicating model attention" (residents: 65\%, senior radiologists: 79\%) were determined as essential factors by both groups. Surprisingly, a majority of senior radiologists (86\%) also selected ``offering Standardized Radiological Terminology and Classification Systems (SRTCS)" as an important factor, whereas no residents selected this option. Other determining factors chosen by both groups included ``generating textual explanations", ``showing representative cases for an outcome". The least selected option was ``indicating counterfactual explanations related to a decision" (33\%).

\subsection{Explainability and Trust}
\label{4.3}

Out of 46 participants, 42 highlighted that explainability is a necessary prerequisite to have trust in ML. Therefore, we investigated different ML model elements which could potentially make it explainable or trustworthy to radiologists. Accordingly, having transparent data preprocessing and curation steps, having transparent model training, e.g., hyperparameter finetuning, and having a feedback mechanism were mostly determined as the aspects enhancing both explainability and trustworthiness. It is also worth noting that being integrated into an existing software used by radiologists, e.g., medical image viewer, was not widely selected as a primary factor for enhancing ML explainability or trustworthiness. 

A key observation was that most participants indicated they would not completely reject an ML model’s diagnosis if it contradicted their own, but would instead seek additional information before making a final decision  (Figure \ref{contradict} in Appendix \ref{appendix2}). Sixteen residents and 10 senior radiologists would analyze the explanations provided by the model. Moreover, in contradiction to the residents, a high number of senior radiologists would also consult with their colleagues or refer to the relevant literature. Only three participants mentioned that they would not trust the model diagnosis at all.

\subsection{Model Evaluation and Deployment}
\label{4.4}

\begin{figure*}
\centering
\includegraphics[width=5in]{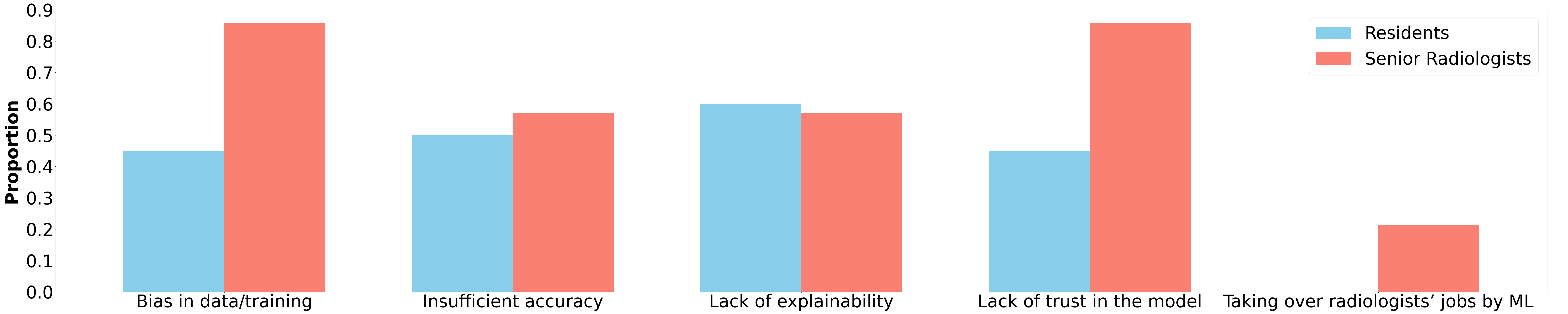} 
\caption{The major concern regarding the use of ML in clinical settings}
\label{concern}
\end{figure*}

The participating radiologists identified ``bias in data/training" (which could affect model fairness), ``lack of trust in the model", ``lack of explainability", and ``insufficient accuracy" as major concerns regarding the use of ML in clinical settings. As seen in Figure \ref{concern}, 86\% of the senior radiologists were concerned about data/training biases, while only 45\% of the residents introduced these biases as a major issue in applying ML to clinical practice. Moreover, as Figure \ref{bias} in Appendix \ref{appendix2} indicates, only 50\% of the senior radiologists think that an explainable ML model can better reveal potential data/model bias(es), with the rest 50\% being unsure about it. This requires further investigation to determine how explainability methods can reveal such biases to senior radiologists more effectively.


These radiologists consider three main factors for evaluating ML outputs, namely the model's task performance, its confidence in the results, and the quality of its explanations in highlighting task-relevant features, in order. Figure \ref{validity} in Appendix \ref{appendix2} demonstrates the proportion of the responses to each of the aforementioned inquiries. Additionally, a majority of the participants specified external (e.g., multi-site) (78\%) and prospective (74\%) validation as crucial aspects of model evaluation.

Approximately 70\% of the radiologists mentioned that an ML model performing similar to radiologists, rather that necessarily outperforming them, can be integrated into the clinical workflow (given that other requirements are satisfied). Unexpectedly, most of them declared that an ML model that consistently performs comparable to radiologists but lacks explainability could still have the potential to be incorporated into their workflow. In addition, a majority of the participants (64\%) confirmed that if a model with radiologist-level performance works independently of radiologist-provided ROIs, taking the whole image as input, they would use it in clinical practice. As another finding, although being integrated into an existing imaging software used by radiologists was not highlighted as a key aspect for improving model explainability and/or trustworthiness, 65\% of the participants considered it as a necessary requirement for using the model in clinical settings.

\section{Discussion} 




The participating radiologists highlighted that they mostly need ML as a helping hand to perform routine clinical tasks, e.g., workflow management and segmentation. This would allow them to dedicate more time to focus on rare and challenging cases. Furthermore, detecting subtle lesions missed by radiologists and flagging urgent cases are other important tasks mentioned by the participants. Additionally, a number of them declared that radiology report preparation is one of the tasks they hope ML can assist in, but they are unsure if it is possible. 

Explainability methods can provide additional support to radiologists by improving their performance in some of the aforementioned tasks. For instance, a visual attention map provided by an ML model for highlighting informative image areas contributing to a diagnostic task can not only be used for image segmentation but also can introduce new biomarkers for the task at hand. As another example, a model that generates Impression and Findings section of radiology reports can be used as a preliminary reference for detecting crucial image areas, including subtle elements, by radiologists for writing these reports. 

Focusing on clinically relevant features has been determined as the most important requirement for ML explainability. As such, a high proportion of the senior radiologists highlighted the importance of explaining ML outcomes based on SRTCS, i.e., radiologist-provided score systems related to a diagnostic task. These findings demonstrate how mapping ML predictions into radiologist-understandable features, which are semantically related to the task at hand, can enhance model explainability.

The participants highlighted several concerns about the deployment of ML in clinical settings. One of most notable ones indicated by most of the senior radiologists is potential biases in data/training. Nevertheless, only half of them agree that explainable ML can reveal these biases. To that end, it is highly necessary to ensure that explainability tools reveal only the actual features applied by a model to make a prediction. Therefore, if there is a spurious correlation among the highlighted features, it can demonstrate a potential source of bias incorporated into model training. However, whether such spurious correlations can indeed be related to data biases requires significant experiments.

Despite gathering insightful findings on the radiologists' viewpoint on explainable ML, there are a number of limitations in this study. First, the number of participants in our survey was low. By collecting more responses, we can enhance the generalizability of our results. Moreover, lack of trust in ML models has been shown to be a major concern among the senior radiologists participating in our survey. Nonetheless, we have not included sufficient questions on trust. Identifying the explicit requirements of trust from radiologists' perspective should be an important future direction. Finally, model confidence in its predictions can be regarded as another useful explainability tool since the participants determined it as an important factor for ML evaluation. However, it was not considered as a major form of explability when designing this survey. Therefore, exploring how this tool can improve ML explainability should be investigated in future work.

In general, the senior radiologists indicated to be more conservative towards the use of ML in  clinical practice. Although addressing the requirements specified by all radiologists is essential, significant effort should be made into satisfying the expectations of senior radiologists as they are more experienced and their trust in ML models should be the first priority in model development. The findings of this study revealed useful insights into the specific needs of this subgroup as well as the general perspective of all radiologists on ML explainability. Following the guidelines proposed in Subsection \ref{4.6} can facilitate model integration into radiological settings, resulting in more efficient clinical workflows.

\subsection{Design Guidelines}
\label{4.6}

Based on the feedback from the participating radiologists, we propose a set of design guidelines in the development and evaluation of explainable ML in radiology across four stages of the ML pipeline: problem framing and data preprocessing, model training, explainability evaluation, and refinement.

\subsubsection{Problem Framing and Data Preprocessing}

The clinical problem should be precisely defined in close consultation with relevant radiologists at early stages. It is important to reassure them that the ML model will serve as a helping hand for increasing their efficiency and that they will still remain at the core of decision-making process. Moreover, training data should be as diverse as possible, covering different patient subgroups and containing multiple modalities. Models trained on such datasets are more likely to generate clinically valid explanations. Exploratory data analysis, e.g., visualizing data distribution and feature correlations prior to model training, can be an effective way of enhancing the transparency of collected data. Furthermore, data preprocessing steps should be clearly illustrated to radiologists, potentially through visualizing the medical images before and after preprocessing.

\subsubsection{Model Training}

Radiologists need to understand the rationale behind specific training configurations or hyperparameter choices. To demonstrate this, ablation studies can be performed by removing individual model elements and reporting the resulting changes in model performance. Providing clear visualizations or simplified illustrations of the training process can further facilitate the comprehension of model design.

Integrating appropriate sources of domain knowledge, e.g., radiology reports, into model training can not only enhance its performance but also make the model focus on clinically relevant image areas for making a prediction. This would align the model decision with clinical evidence, as one of the most essential explainability requirements indicated by the participants. To that end, multimodal foundational frameworks trained on large training data consisting of different modalities, e.g., medical images and radiology reports, have higher potential to encode relevant semantic information, thereby focusing on radiologist-understandable features. Because they are trained on large amounts of data, these models can generalize to different settings and downstream tasks. Moreover, the ability to generate multiple explanation forms, including visual heatmaps and textual reasoning, makes them more suitable to satisfy radiologists' requirements for explainability. 

With respect to explanation methods, post-hoc approaches such as Grad-CAM are not highly reliable, as they rely on external algorithms to identify informative features. On the other hand, integrating explanations into model training using self-explainable methods (e.g., trainable attention) could be more valid as these attention weights are learnt during training, thereby alleviating the need for applying external algorithms to compute the weights \citep{hou2024self}. Nevertheless, the degree to which these explanations reveal model decisions still requires further investigation \citep{wiegreffe2019attention}.

SRTCS, i.e., radiologist-provided scoring systems corresponding to a diagnostic task, has been identified as an essential explainability tool among the participating senior radiologists. 
Generating SRTCS scores along with other explanation forms, such as visual heatmaps, can provide a higher level of trust among senior radiologists, helping them map the model output to their semantic measurements. Providing SRTCS scores is possible using different methods. For example, an efficient method is to generate SRTCS scores as part of textual explanations within radiology reports. This method involves identifying and generating relevant semantic features for a given case, followed by determining the corresponding assessment based on those features. Another method is to incorporate SRTCS scores as an additional data modality during training and add an additional module to the model architecture for encoding them.

\subsubsection{Explainability Evaluation}
\label{eval}

Based on the survey findings and previous works in the literature, we have identified the following metrics for assessing an ML explainability tool:

\begin{itemize}
    
    \item Alignment with clinical evidence: The features or image areas specified by an explanation should be semantically related to the task at hand and align with radiologist knowledge. To that end, the overlap between explanation-highlighted features and radiologist-provided information, e.g., manual segmentation masks, should be provided to enhance the reliability of the explanation.
    \item Fidelity: An explanation should accurately reveal features determining in ML predictions. For instance, a visual heatmap should highlight only image areas attended by the model in the decision-making process. This could be measured by perturbing a feature highlighted by an explanation and monitoring the impact on model performance. If the impact is significant, the highlighted feature can be highly determining in model predictions, confirming the fidelity of the explanation to the model decisions.

    Providing this metric for an explanation could ensure that the explanation truly reflects any semantic or spurious features contributing to ML outcomes. Spurious correlations can correspond to potential data biases \citep{hamidieh2024views,ahmadi2023mitigating}. Therefore, showing that such correlations have high fidelity to ML outcomes can partially address the major concern of the participating senior radiologists, i.e., data/training biases, as discussed in Subsection \ref{4.4}.
    
    \item Robustness: There are a wide range of variations across different radiological settings, including scanner shifts. A valid explanation should be robust to these variations and remain consistent for the same task across different workflows. Nevertheless, indicating this can be challenging given the high number of such variations. A practical method could be adding adversarial noise to training data, which would simulate these alterations, and assessing its influence on an explanation. Insignificant changes in the explanation can demonstrate a degree of robustness.
    
    \item Effect on radiologist performance: The participants in this survey stated that they would mainly prefer ML to be their helping hand in routine clinical tasks. They also indicated that explainable ML can signify useful semantic features for a diagnostic task and assist them in writing radiology reports. Therefore, explainability assessment should consider whether the model operates in alignment with the requirements of its end users (radiologists), e.g., detecting accurate imaging features to be included in radiology reports. Consequently, the effectiveness of the explanation can be determined by measuring its impact on the user's performance in the clinical task at hand. 
    
\end{itemize}

\subsubsection{Model/Explanation Refinement}

Designers should treat explainability as an iterative refinement process. Based on the evaluation metrics illustrated in Subsection \ref{eval}, ML explanation tools may require multiple adjustments to ensure that they meet necessary criteria for clinical deployment. These adjustments include, but are not limited to, maximizing the overlap between model attention maps and manual segmentation or increasing the consistency of explanation tools when different noise functions are applied to training data. Such adjustments can be operationalization by adding additional components to training objectives or loss function used for training the model, encouraging it to generate desirable and clinically meaningful explanations.

Moreover, interactive feedback loops can be integrated into model design through which radiologists could review and revise model predictions if necessary. These revisions could then be incorporated back into the model to improve its performance over time. This feedback loop would also allow radiologists to investigate and comment on different explanations generated by the model, shaping the explanatory behavior of the model based on clinical needs and reasoning.

\section{Conclusion}

In this study, we surveyed a group of radiologists with varying experience levels from different hospitals in a metropolitan area to gather their perspectives on ML model explainability and identify the most effective explainability tools for justifying model decisions. These radiologists defined explainability as an ML model's ability to focus on clinically relevant imaging features and generate multiple forms of explanations including SRTCS scores, visual heatmaps, and textual reasoning. Moreover, data/training biases has been highlighted as a primary concern among the participating senior radiologists, and the potential of ML explainability methods in revealing such biases is a crucial issue which should be addressed in future work. We also developed a set of guidelines based on the survey results to help ML researchers design more effective explainability methods, accelerating model integration into the clinical workflow.

\section{Funding Acknowledgment}
This work was supported by Natural Sciences and Engineering Research Council of Canada (NSERC).

\bibliography{chil-sample}

\appendix

\section{Existing Design Gaps}\label{appendix1}

\label{4.5}

In this subsection, we highlight the gaps in the existing literature based on the findings of this survey. First, clinical tasks for ML applications should be identified through close consultation with clinicians. Many works in the literature have explored tasks that may not be highly useful to radiologists, and therefore do not meaningfully enhance their clinical workflow. Rather than aiming to replace radiologists in a single aspect of their work, ML models should be developed as supportive tools to improve their efficiency across multiple tasks.

Next, according to the survey results, providing explanations only for model outcomes is not enough, and a comprehensive investigation of data sources, preprocessing steps, and model configuration is also required to enhance model explainability. To the best of our knowledge, many proposed explainability approaches focus solely on model outcomes, without uncovering the mechanism of other model development stages. Clarifying the rationality behind all model development steps can also be insightful for showing potential data biases, which has been identified as a major concern among the participating senior radiologists.

As another notable finding of this survey, ``SRTCS'', i.e., radiologist-provided diagnostic scoring systems, has been widely highlighted by the participating senior radiologists as an essential factor for ensuring model explainability. However, to the best of our knowledge, SRTCS has been investigated as an explanation tool for justifying model outcomes only in a few works \citep{hamm2023interactive}. Moreover, as indicated by the participants, other explanation tools, including visual heatmaps and textual explanations, are also required to verify model decisions. Uni-modal ML models, i.e., those taking only one modality (e.g., image) as input, which are the core of many existing works, may not fulfill this requirement, and the development of multimodal ML frameworks providing multiple explanation forms should be explored. 

In addition to understanding model decision-making process, this study revealed other important reasons for prioritizing model explainability. These include verifying the clinical relevance of features that are highly correlated with ML outputs as well as increasing radiologists' confidence in writing radiology reports. Therefore, Effective explainability approaches can help radiologists identify informative biomarkers for a diagnostic task, including subtle findings they might otherwise miss. Furthermore, as specified by the senior radiologists, bias in data/training is one of the major issues preventing ML models from being adopted in clinical settings, and almost half of them were not certain if ML explainability can demonstrate these biases. This could be one of the essential factors hindering trust in ML among radiologists. Thus, substantial research is needed to determine whether explainability methods can uncover such biases and effectively communicate them to radiologists.



Finally, ML evaluation should involve model performance, explainability tools, and uncertainty measurements. Specifying uncertainty measurement gaps is outside the scope of this work, and we only explore explainability evaluation. 
To that end, 
The provided explanations should align with clinical knowledge and accurately reflect the features driving model predictions. 
Post-hoc explainability methods, which have been intensively explored in the literature, cannot typically meet this requirement as their fidelity to model decision is unclear. In addition, the fidelity of self-explainable methods, such as trainable attention-based methods, to ML model outcomes needs further investigation, as emphasized in \citep{wiegreffe2019attention}. 

\newpage
\section{Supporting Figures}\label{appendix2}

\begin{figure*}[b]
\centering
\includegraphics[width=6.1in]{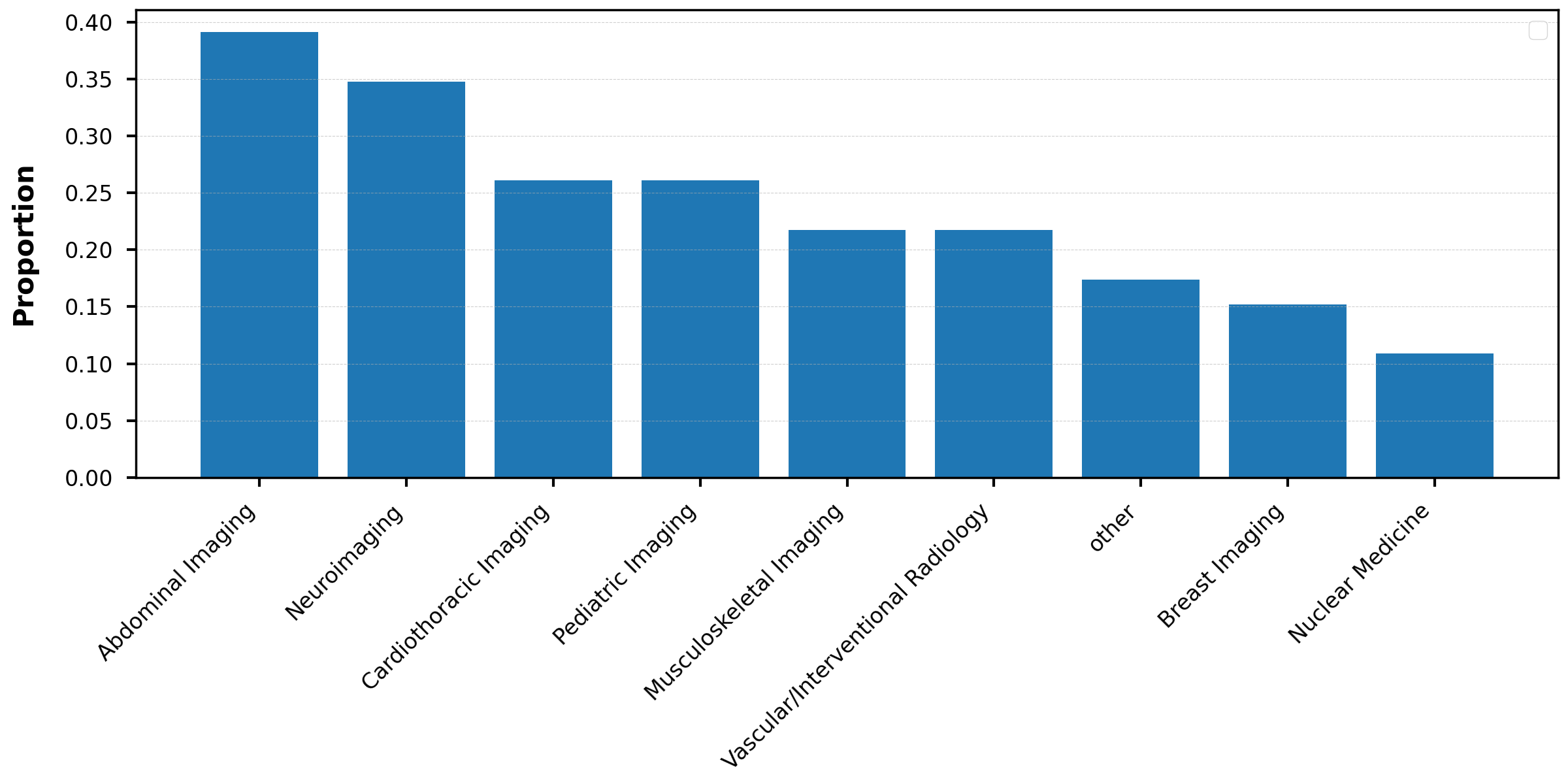} 
\caption{The Main Clinical/Radiological Focus of the Survey Participants}
\label{focus}
\end{figure*} 
\vspace{-9in}
\begin{figure*}[ht!]
\centering
\includegraphics[width=6in]{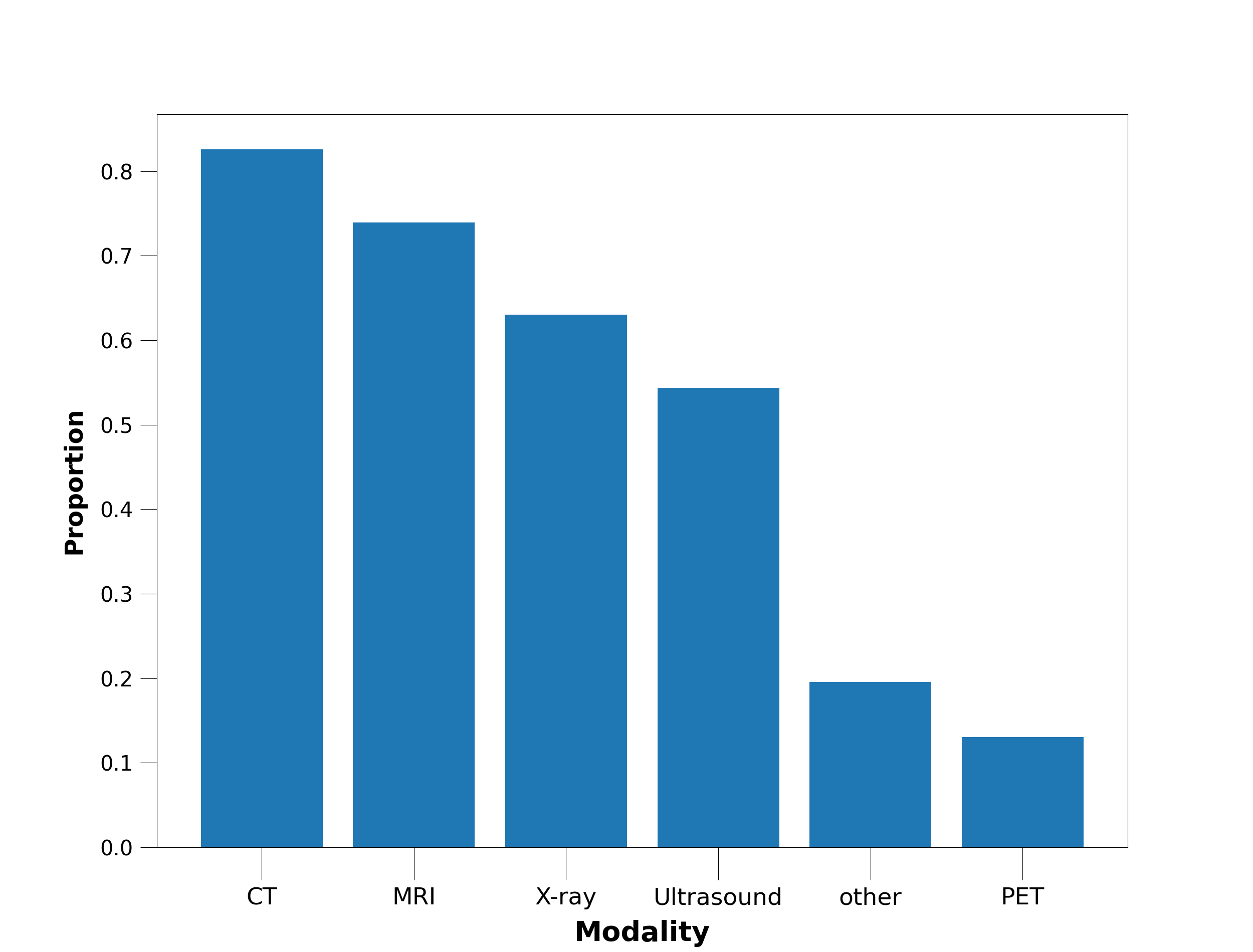} 
\caption{The Imaging Modality Used in the Participants' Main Clinical Focus}
\label{modality}
\end{figure*}

\begin{figure*}[ht!]
\centering
\includegraphics[width=6in]{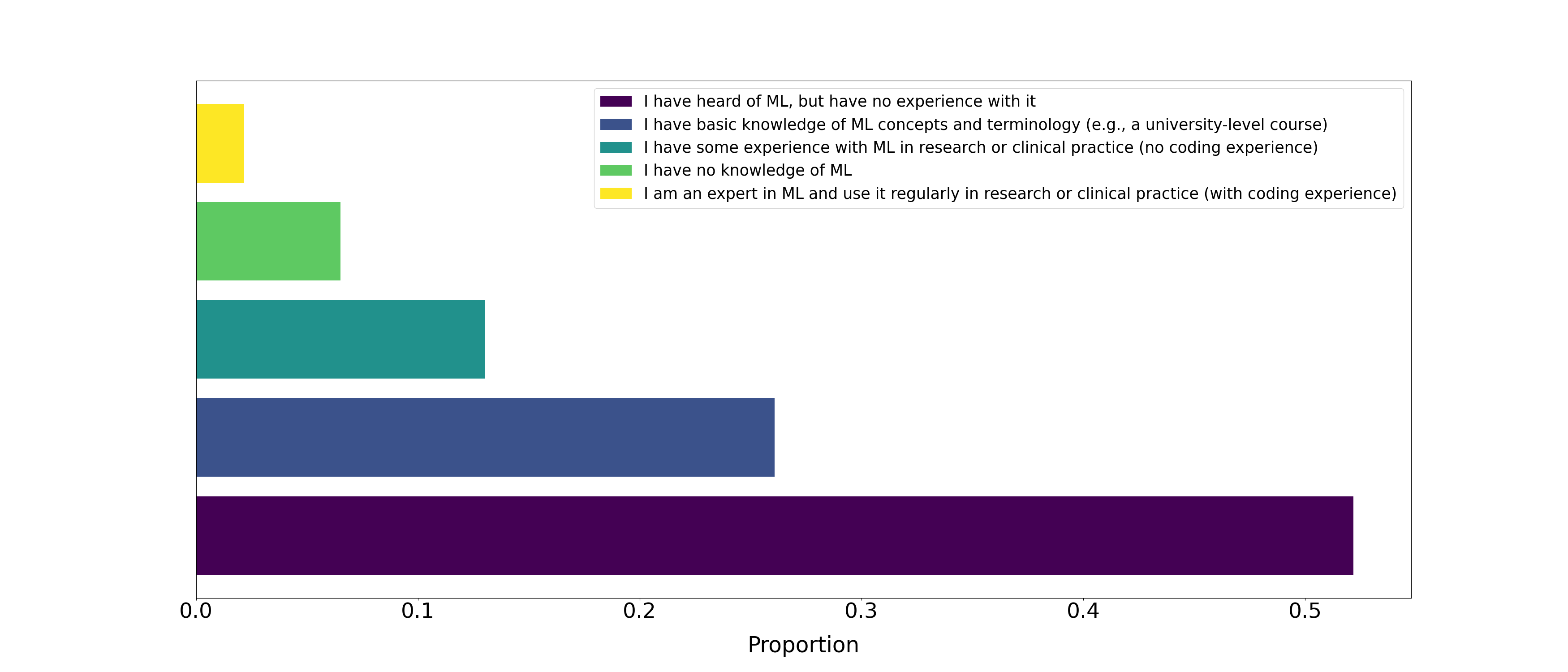} 
\caption{The Participants' Familiarity with ML}
\label{ML}
\end{figure*} 

\begin{figure*}[ht!]
\centering
\includegraphics[width=6in]{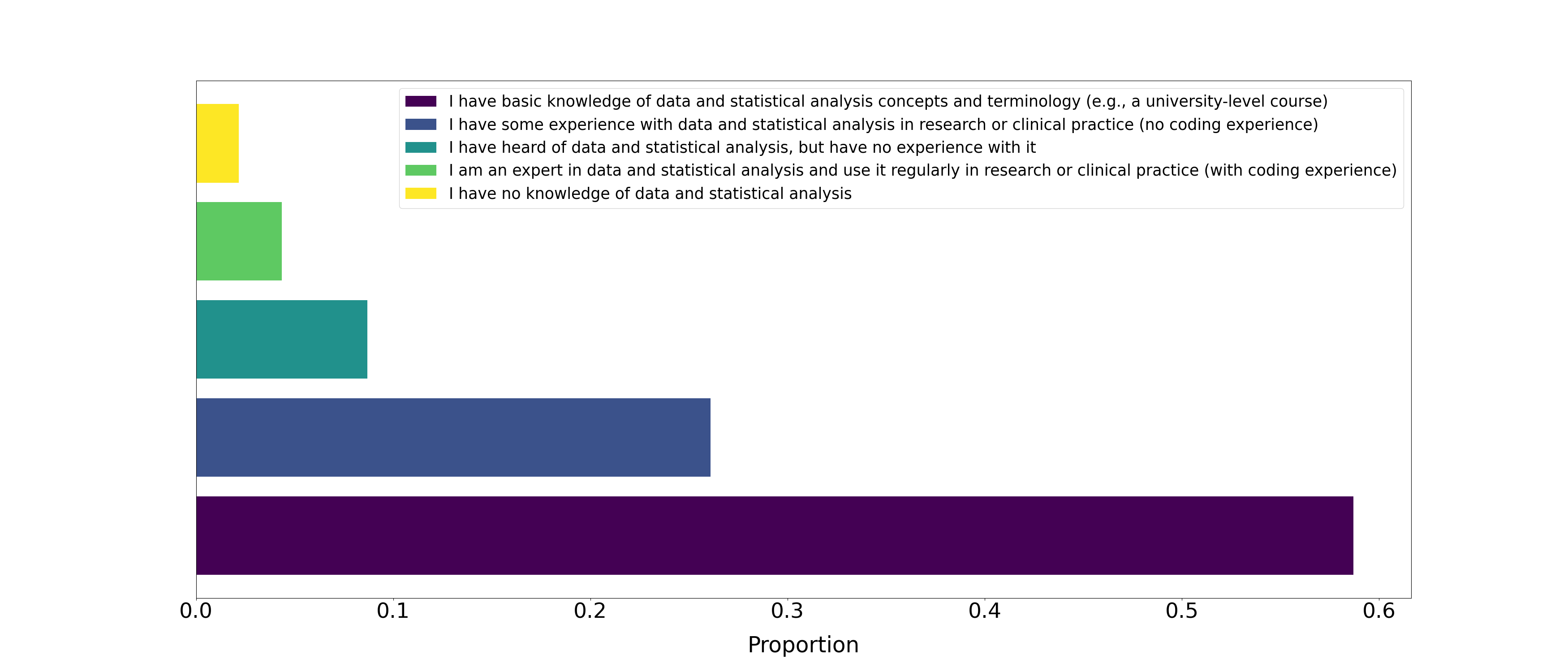} 
\caption{The Participants' Familiarity with Data and Statistical Analysis}
\label{stat}
\end{figure*}

\begin{figure*}[ht!]
\centering
\includegraphics[width=6.6in]{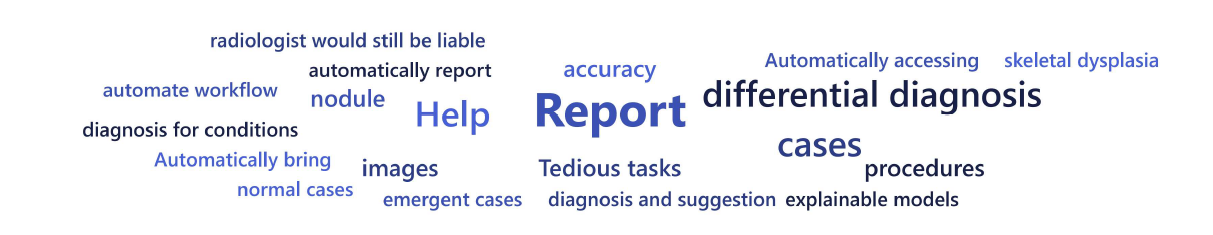} 
\caption{Potential Applications of ML in Clinical Practices From the Participants' Perspective}
\label{open_ended_2}
\end{figure*}

\begin{figure*}[ht!]
\centering
\includegraphics[width=6in]{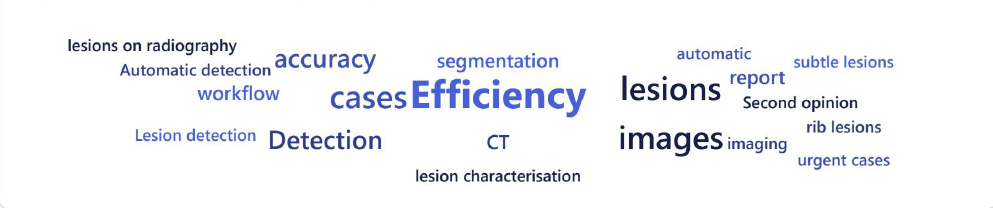} 
\caption{Useful Applications of ML in Clinical Practices From the Participants' Perspective}
\label{open_ended}
\end{figure*}

\begin{figure*}[ht!]
\centering
\includegraphics[width=5in,height=3in]{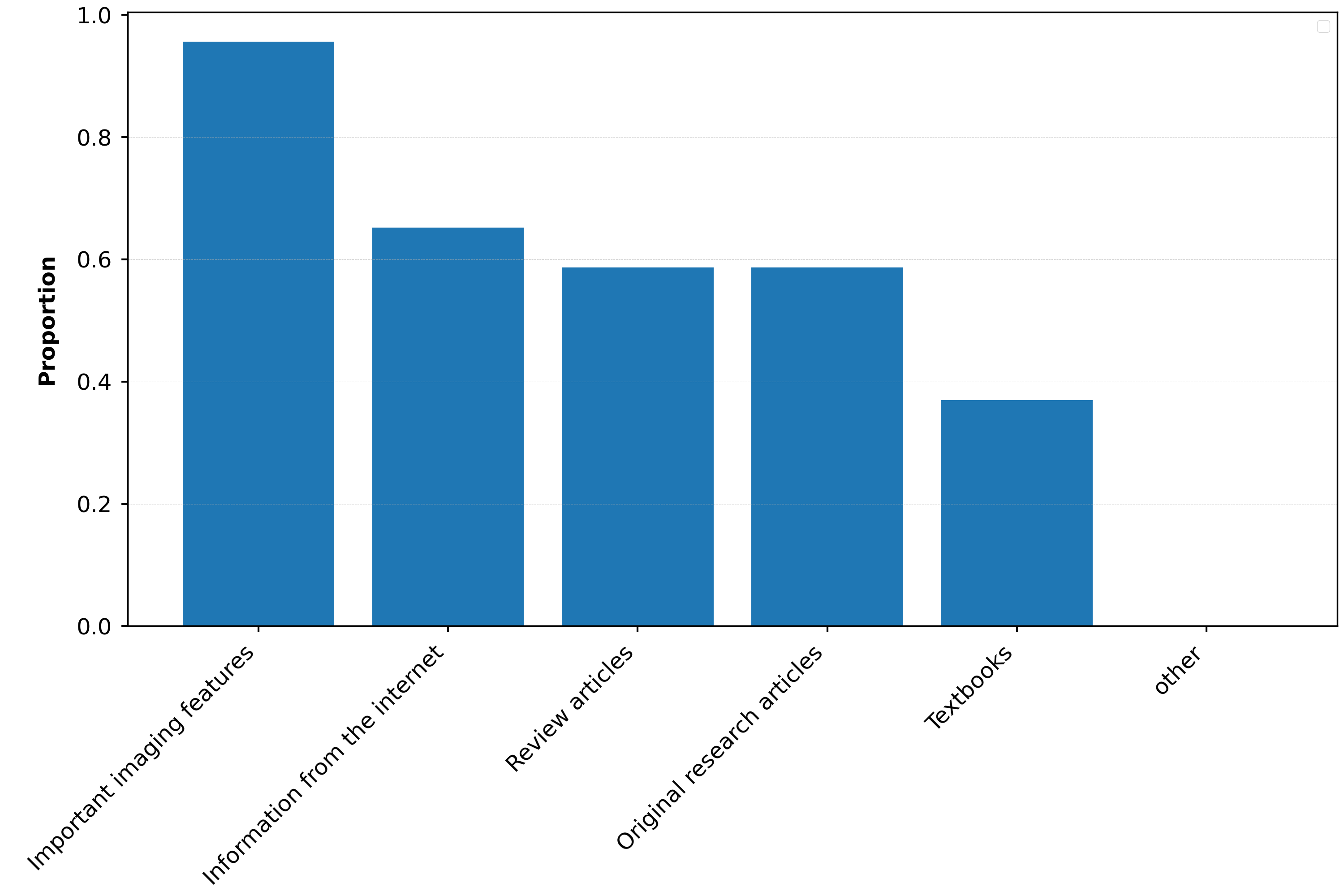} 
\caption{Explanation Tools Used by Radiologists}
\label{tools}
\end{figure*}

\begin{figure*}
\centering
\includegraphics[width=6.6in]{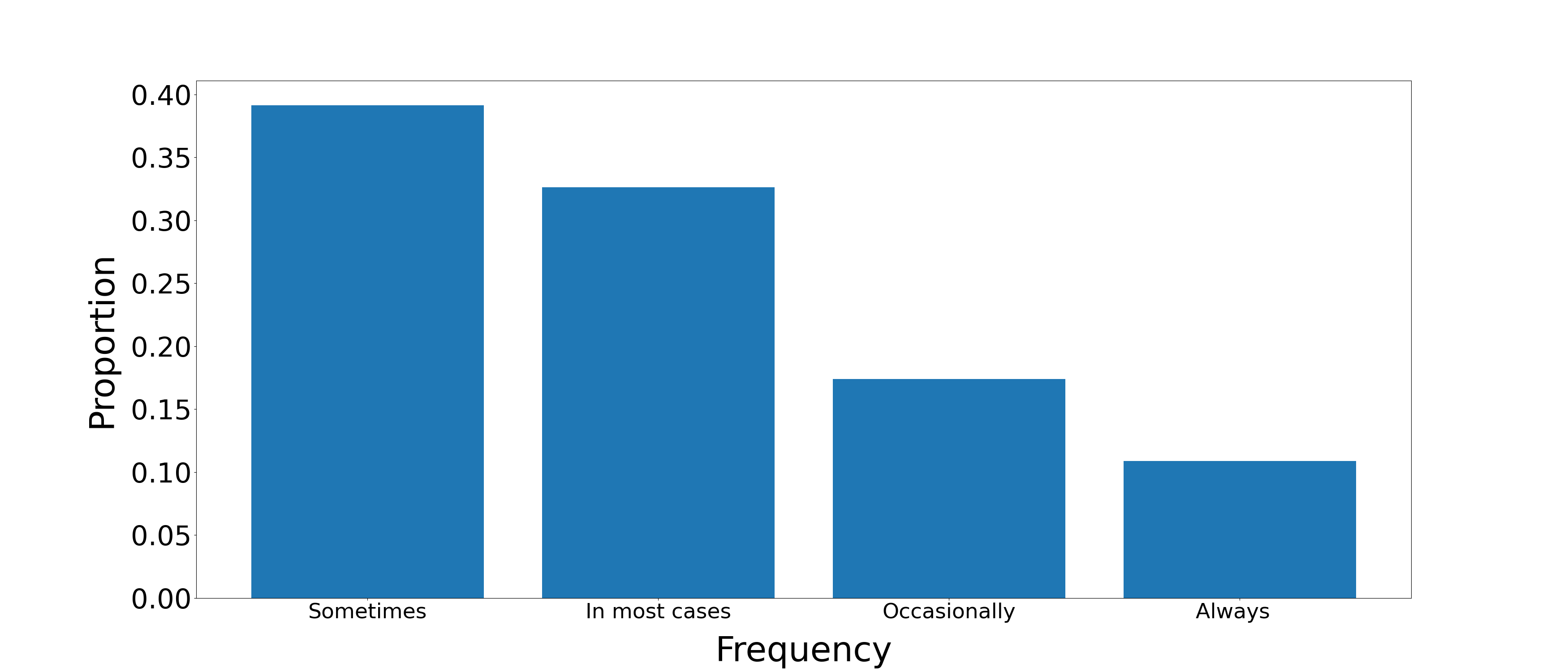} 
\caption{The frequency of Using Similar Cases in the Radiologists' Workflow}
\label{similar}
\end{figure*} 

\begin{figure*}[ht!]
\centering
\includegraphics[width=6.1in]{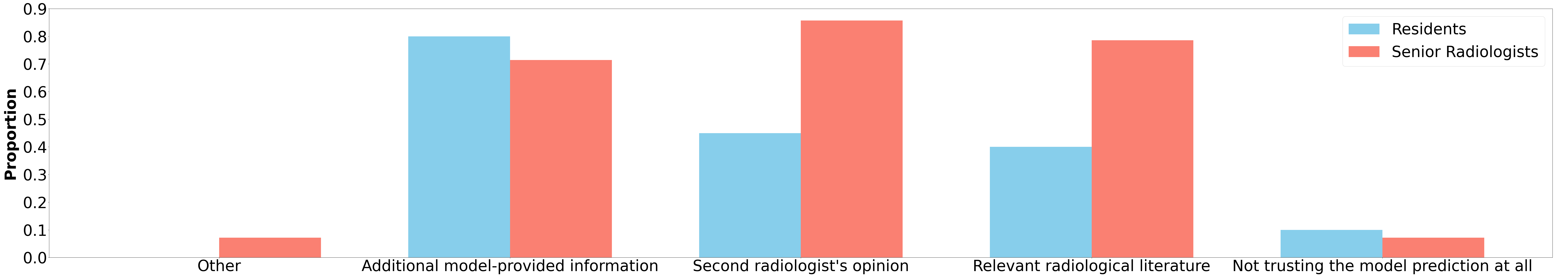} 
\caption{What the participants would use/do if ML diagnosis contradicts theirs}
\label{contradict}
\end{figure*} 

\begin{figure*}[ht!]
\centering
\includegraphics[width=2.5in]{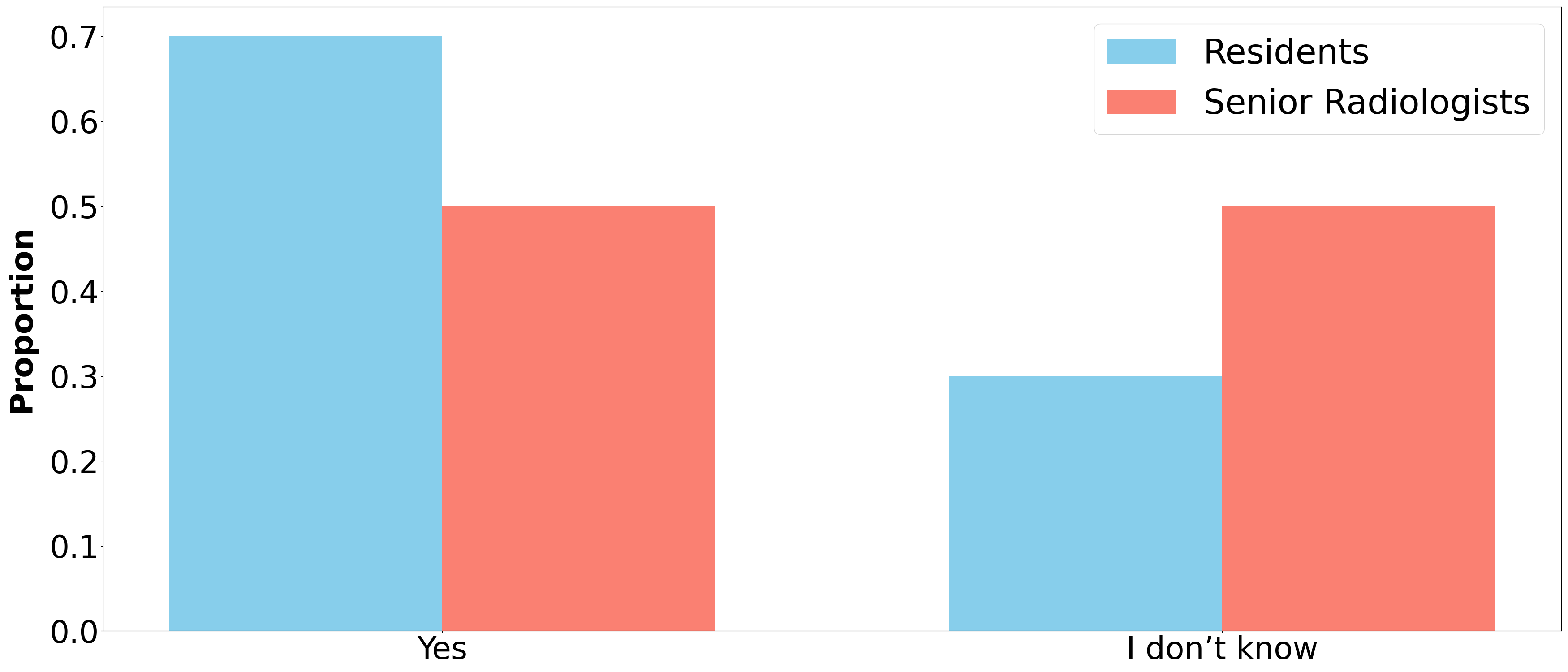} 
\caption{Whether explainable ML can better reveal potential data/training biases}
\label{bias}
\end{figure*} 

\begin{figure*}[ht!]
\centering
\includegraphics[width=3.5in]{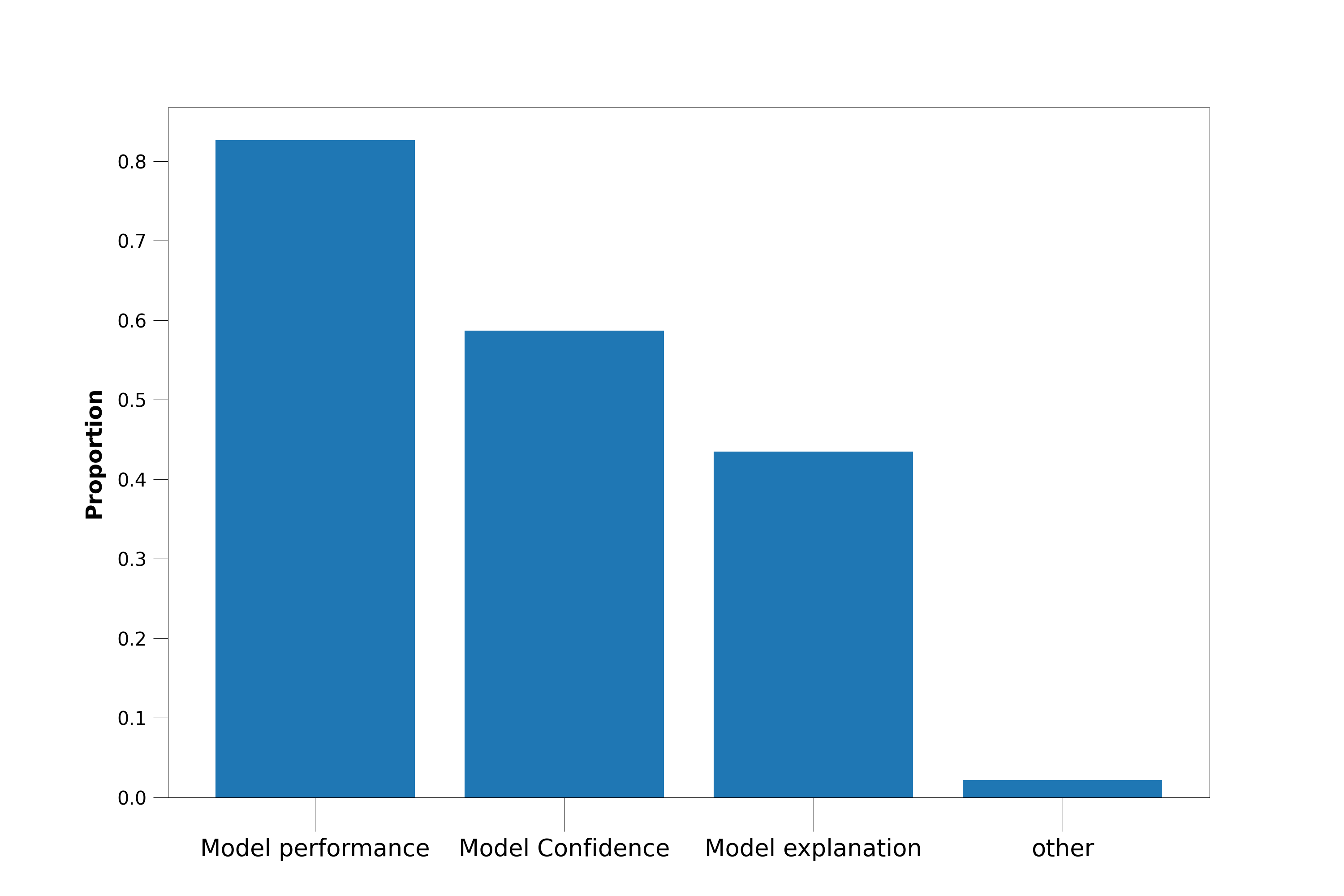} 
\caption{ML evaluation aspects considered by the participants}
\label{validity}
\end{figure*}

\end{document}